\documentclass[showpacs,pra,aps,superscriptaddress,twocolumn,floatfix]{revtex4}
\usepackage{graphicx,float,amsmath,amssymb,mathrsfs}
\def\sss{\scriptscriptstyle}

\begin{document}

\title{Breakdown of superfluidity of a matter wave in a random environment}

\author{M. Albert}
\affiliation{ D\'epartement de Physique Th\'eorique, Universit\'e de 
Gen\`eve, CH-1211 Gen\`eve, Switzerland}
\author{T. Paul} \author{N. Pavloff} \author{P. Leboeuf}
\affiliation{Univ. Paris Sud, CNRS, Laboratoire de Physique
  Th\'eorique et Mod\`eles Statistiques, UMR8626, F-91405 Orsay}

\begin{abstract}
  We consider a guided Bose-Einstein matter wave flowing through a disordered
  potential. We determine the critical velocity at which superfluidity
  is broken and compute its statistical properties. They are
  shown to be connected to extreme values of the random potential.
  Experimental implementations of this physics are discussed.
\end{abstract}

\pacs {03.75.Kk~; 05.60.Gg~; 67.10.Jn}

\maketitle

\section{Introduction}
The simplest and most intuitive definition of superfluidity (SF) is the
ability to move without dissipation. According to a perturbative mechanism
proposed by Landau, superfluidity is broken 
 when the velocity of the flow exceeds a critical value $V_c^{\rm
  L}$ at which it is energetically favorable to emit elementary excitations.
Though this mechanism has been explicitly verified in $^4$He
\cite{All76}, $^3$He-B \cite{Cast86} and Bose Einstein condensates
(BECs) \cite{Chik00}, many experiments in $^4$He \cite{Ave85},
$^3$He-A \cite{3hea} and BECs \cite{Raman99,Engels07} have shown
that the actual critical velocity $V_c$ is generally lower than
$V_c^{\rm L}$ due to the occurrence of phase slips.
In this scenario, SF is protected by an energy barrier which may be
overcome by fluctuations of thermal (as first suggested by Iordanskii in the
context of liquid He-II \cite{Ior65} and by Little for superconductors
\cite{Lit67}) or quantal origin (as more recently observed in $^4$He
\cite{quantum4he}, in superconducting nano-wires \cite{Lau01} and possibly
also in BECs \cite{McK08}), leading to what is called a resistive state in the
physics of superconductors.

In what follows we address the problem of determining the critical
velocity for the breakdown of SF of a matter wave moving in a
disordered potential. The matter wave beam is formed by a guided BEC
in a quasi one-dimensional (1D) geometry. We assume zero temperature;
it is well known that in this case superconductivity and superfluidity
are not destroyed by weak disorder. This is Anderson's theorem for
non-magnetic impurities in superconductors \cite{And59}; similar
results (with a different physical mechanism) hold for BECs
\cite{weakdis}. In the latter case, the phase coherence of the system
is preserved in 1D in the presence of a weak disorder as demonstrated
in the experiments reported in Refs. \cite{Cle88,Chen08}. As a
consequence, one may study a simple scenario for breakdown of
superfluidity in disordered BECs where phase slips are neither
thermally nor quantum mechanically nucleated but rather have a
dynamical origin : the barrier disappears at a given critical
velocity. This mechanism is standard in the absence of disorder (see,
e.g., \cite{Stri09} and references therein) and the extraordinary
control achieved in the domain of atomic vapor has even allowed a
direct observation of the nonlinear excitations nucleated above the
critical velocity $V_c$ \cite{Raman99,Engels07}. However, to our
knowledge there is up to now only one clear experimental evidence of
dynamical breakdown of SF and of finite critical velocity in the
presence of disorder, obtained by studying the damping of dipole
oscillations in an elongated BEC \cite{Hulet09}.  In our fully 1D
case, as well as in the dipole oscillation experiments \cite{Hulet09},
an important issue is to understand the out-of equilibrium solutions
of a nonlinear continuous system in the presence of disorder. In this
context, the phase diagram of the fluid flowing through a quasi 1D
disordered potential $U(x)$ of finite extent $L$ was recently studied
in Refs. \cite{Paul07,Alb08,Paul09}. Here, we concentrate on the SF
part of this diagram and more specifically on the description of the
breakdown of SF when the velocity (or the length $L$ of the disordered
region) increases.

We study two different types of disordered potential with opposite
characteristics. The first one is a smooth potential whose typical
spatial scale of variation is large compared to the healing length of
the condensate. In this case a local density approximation holds and a
local Landau criterion can be applied \cite{Hakim97}. This mechanism
reconciles the Landau approach with the phase slip phenomenon,
because it predicts that SF is broken when the local Landau velocity
is reached by emission of nonlinear excitations (solitons in our 1D
case). The second type of disordered potential consists in a series of
point-like impurities and thus has, contrarily to the previous type,
strong fluctuations on small spatial scales. We show that in this case
the criterion can be adapted yielding -- as in the previous case --
very good agreement with numerical simulations.  In both cases we
explicitly compute the statistical properties of the critical velocity
$V_c$ and show that they are closely related to the extreme value
statistics of the disordered potential. We finally discuss
experimental realizations of our models.

The system considered is a weakly interacting BEC transversely
confined by a harmonic potential of frequency $\omega_{\perp}$. For
simplicity, the disordered potential $U$ is supposed to depend on a
single spatial variable -- the coordinate $x$ along the axial
direction of the guide. 
A stationary flow of the system is then accurately described by a
1D order parameter $\psi(x)$
obeying the nonlinear Schr\"odinger equation \cite{Jack98,Leb01}
\begin{equation} \label{10}
\mu\, \psi=
-\frac{\hbar^2}{2m}\frac{d^2\psi}{d x^2} + \left[
 U(x)+g\, n^{\nu}(x) \right]\psi \; .
\end{equation}
Here, $n(x)\equiv |\psi(x)|^2$ is the condensate density per unit of
longitudinal length, $\mu$ is the chemical potential and
$g=2\hbar\omega_\perp a^\nu$ is the nonlinear parameter ($a>0$ is the
$3$D s-wave scattering length).  In the low density regime ($a n\ll
1$) the density profile in the transverse direction is Gaussian-shaped
and $\nu=1$, whereas $\nu=1/2$ in the opposite high density regime ($a
n\gg 1$) where the Thomas-Fermi approximation holds for the transverse
degree of freedom \cite{rem1}.

\section{Superfluid flows} 
In addition to the density $n(x)$, it is convenient to characterize
the flow by its velocity $v(x)=\frac{\hbar}{m}
\,[\mbox{arg}\,(\psi)]_x$.
We assume that the disordered potential $U(x)$ takes sizable values only
over a region of finite length $L$. In this case a SF flow corresponds to a
solution of Eq. (\ref{10}) with constant density $n_0$ and velocity
$V$ at $\pm\infty$. The chemical potential $\mu$ then reads:
\begin{equation}\label{mu}
  \mu=\frac{1}{2}m V^2+g\,n_0^\nu \; .
\end{equation}
In the following we denote the chemical
potential of a BEC at rest as $\mu_0$ ($\mu_0=g\,n_0^\nu$).

In absence of external potential, the SF solution corresponds to
$n(x)=n_0$ and $v(x)=V$ for all $x$, and one can show that it is stable
under a weak perturbing potential provided $V\leq c_0$ where
\begin{equation}
c_0=\sqrt{\frac{\nu\,g\,n_0^\nu}{m}} 
  \label{sound}
\end{equation}
is the sound velocity of the unperturbed condensate. The condition
$V\leq c_0$ is exactly the Landau criterion for SF
because in a BEC the velocity $V_c^{\rm L}$ is precisely the
speed of sound.

\section{Slowly varying disordered potentials} 
In the case of a slowly varying potential, that is, when the typical length of
spatial variations of $U(x)$ is much larger than the healing length
$\xi=\hbar/\sqrt{m\mu_0}$ of the fluid, one can devise a local density
approximation for describing stationary flows. In this scheme, the
flow verifies current conservation and local equilibrium. This reads
$n(x)v(x)=C^{\rm st}=n_0 V$ and
\begin{equation}\label{le}
 \mu=\frac{1}{2}mv^2(x)+ g\,n^\nu(x)+U(x)\; .
\end{equation}
From these considerations it is easy to see that the velocity $v(x)$
is determined by a simple algebraic equation \cite{Hakim97}:
$U(x)=\mu_0\,G(v(x))$ where $G(v)=
1-(V/v)^\nu+\frac{\nu}{2}(V^2-v^2)/c_0^2$. This equation admits a
solution provided $U(x)/\mu_0$ is lower than the maximum of $G$. If
this condition is violated the flow no longer admits a stationary
solution, SF is broken and the flow becomes dissipative (one can show
that in this case the obstacle described by the potential $U(x)$
experiences a finite drag \cite{Pav02}).  The maximum of $G$ is
reached when $V^\nu c_0^2 =[v(x)]^{2+\nu}$, which precisely reads
$v(x)=c(x)$ where $c(x)=\sqrt{\nu g \, n^\nu (x)/m}$ can be termed the
local sound velocity [compare with (\ref{sound})]~: SF is broken when
one reaches the local Landau criterion.

In this approximation, it is clear that SF will first break
down at the point $x=x_{\rm m}$ where the potential is maximum:
$U(x_{\rm m})=\mbox{max}\,[U(x)]=U_{\rm m}$. The condition $v(x_{\rm m}) =
c(x_{\rm m})$ yields an explicit relation between $U_{\rm m}$ and $V_c$:
\begin{equation}\label{vc0}
\frac{U_{\rm m}}{\mu_0}
=1+\frac{\nu}{2}\left(\frac{V_c}{c_0}\right)^2-\left(1+\frac{\nu}{2}\right)
\left(\frac{V_c}{c_0}\right)^{2\nu/(\nu+2)} \; .
\end{equation}

Everything now boils down to a problem of extreme value: once the statistical
properties of the maximum $U_{\rm m}$ of $U(x)$ over $[0,L]$ are known, the
distribution of the critical velocities $V_c$ can be obtained readily through
(\ref{vc0}). Namely, if one denotes by ${\mathscr P}_{\sss\!\!
  L}(U_{\rm m})$ the probability
distribution of $U_{\rm m}$ and by $P_{L} (V_c)$ the corresponding
distribution of the critical velocity $V_c$ one has
\begin{equation}
P_{L}(V_c)= \frac{\nu\mu_0}{c_0} 
\left[ \frac{V_c}{c_0} - 
\left(\frac{V_c}{c_0}\right)^{(\nu-2)/(\nu+2)}\right] 
\, {\mathscr P}_{\sss\!\!  L}(U_{\rm m}) \; .
  \label{dist_vc0}
\end{equation}
The distribution function of $U_{\rm m}$ (and thus also that of $V_c$)
depends on the size $L$ of the disordered region for the simple reason
that the longer the disordered region, the larger the probability of
finding a large maximum of $U$. Hence it is clear that, on average, the
critical velocity decreases with increasing sample length $L$. Note also
that, in this picture, the critical velocity is related to the
local fluctuations of the disorder and not to the details of the
correlations.

In order to calculate the distribution of $U_{\rm m}$, the first step
consists in mapping the problem of finding the maximum value of a
correlated continuous function to a problem of a set of $N$ discrete
uncorrelated variables.  According to \cite{Pickands} this mapping can
be done if the correlation function of $U(x)$, characterized by a
correlation length $\ell_c$, decays faster than a logarithm and
provided $L/\ell_c \gg 1$.  From now on we assume that these
requirements are fulfilled, and study the extreme value statistics of
a set of $N=\gamma L/\ell_c$ uncorrelated random variables
$\{U_1,U_2,...,U_N\}$ dis\-tri\-bu\-ted according to the probability
distribution of the disorder potential $p(U)$. $\gamma$ is a parameter
of order unity that depends on the correlation function and has to be
determined numerically \cite{rem2}.  If we denote by $f(U)$ and
${\mathscr F}_{\sss\!\!  L} (U_{\rm m})$ the cumulative distribution
functions of $U$ and $U_{\rm m}$, respectively, we have ${\mathscr
  F}_{\sss\!\!  L} (U_{\rm m})=[f(U_{\rm m})]^N$. Taking the
derivative of this expression yields the probability distribution
${\mathscr P}_{\sss\!\!  L}(U_{\rm m})$ and then the distribution of
the critical velocity through (\ref{vc0}) and (\ref{dist_vc0}).

To estimate the frontier between the superfluid and the dissipative
regimes, we calculate the median $L(V_c)$ of this distribution and obtain
\begin{equation}
  \frac{L(V_c)}{\ell_c}=\frac{\gamma^{-1}\ln 1/2}{\ln
    f(U_{\rm m}(V_c))}\; ,
  \label{bound}
\end{equation}
where $U_{\rm m}(V_c)$ is given by (\ref{vc0}). 
\begin{figure}
  \includegraphics[width=0.95\linewidth]{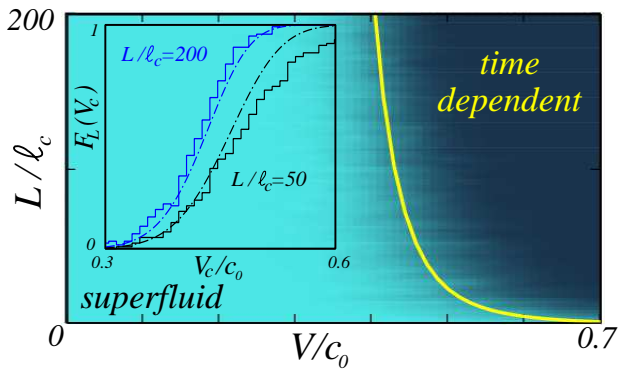}
  \caption{(Color online) Transport of a quasi 1D BEC with velocity $V$ through
  a Gaussian correlated
disordered potential of extension $L$ (the parameters $\nu$,
  $\ell_c$ and $\Sigma$ take here the values $\nu=1$, $\ell_c=5\xi$ and
    $\Sigma=0.1\mu_0$). Dark region: time dependent
  flow; light gray (light blue online) region: SF stationary flow. The
  (yellow online) solid line displays the boundary between the two regions as
  predicted by Eq. (\ref{bound}).
The inset displays the cumulative probability distribution of the
critical velocity $V_c$ for samples of two different lengths. Staircase 
functions: numerical computations; dot-dashed lines: theory.}
  \label{diag_gauss}
\end{figure}

In order to check the validity of our approach, we have numerically
determined the critical velocity from time dependent simulations of
the Gross-Pitaevskii equation. Starting from the ground state in the
presence of disorder at zero velocity, we have adiabatically
accelerated the disordered potential until it reaches a velocity $V$.
For each $V$ and $L$ we consider 80 realizations of the random potential
and determine the fraction $P_s$ of stationary
solutions. This quantity is plotted in Fig.~\ref{diag_gauss} using a
gray scale [dark, $P_s=0$ ; light blue (gray), $P_s=1$] as a function of
the normalized variables $L/\ell_c$ and $V/c_0$.
The inset displays $F_L (V_c)$, the cumulative probability distribution of the
critical velocity. Very good agreement is observed in the
expected limit of validity of our approach ($L/\ell_c \gg 1$). 
Figure \ref{diag_gauss} is drawn in the case of a Gaussian disorder,
$p(U)=\exp(-U^2/2\Sigma^2)/\sqrt{2\pi\Sigma^2}$, with a Gaussian correlation
function (numerically we find $\gamma\simeq 0.8$). We have also checked the
accuracy of our predictions for other types of disorder of
experimental interest (Lorentz-correlated disorder and speckle potential).

\section{A series of $\delta$ scatterers} 
Another commonly used model of disorder is a potential formed by 
a series of $\delta$-like
impurities:
$U(x)=\lambda\mu_0\xi\sum_{i=1}^{N}\delta(x-x_i)$, where the $x_i$'s are
uncorrelated random variables distributed between $0$ and $L$ with
density $\rho=N/L$. 
$\lambda>0$ is the dimensionless strength of a scatterer.  In the
presence of such a potential, the local Landau criterion is no longer
applicable because the density is not smooth. 
However one can devise an approach adapted to this particular case.
One first remarks that each impurity repels the condensate, the
density of which reaches its lowest local value at the position of the
$\delta$ peak. The region in space where the decrease in
density is the largest will correspond to configurations where
two (or more) scatterers lie very close to each other. SF will break
down, therefore, at the point where the local concentration of $\delta$
peaks is maximum.

The smallest length scale for density modulations of the condensate is
the healing length $\xi$. The condensate is therefore not sensitive
to details of the disordered potential on scales smaller than
$\xi$. One thus divides the disordered region into $B=L/\xi$ boxes, each of
size $\xi$, and replaces the $m_i$ delta peaks present in box $i$ by a
single effective peak of strength $m_i\,\lambda$. The goal is then to
determine the probability distribution of intensity of the strongest
of the effective peaks, because it is at this peak that SF will first
be broken. One thus needs to calculate the probability distribution of
$M=\textrm{max} \left\{{m}_1,{m}_2,...,{m}_B\right\}$, where $\sum_i
{m}_i = N$.

The probability of finding ${m}$ peaks in an interval of length $\xi$
is $\pi({m})= \binom{N}{m} p^{m} (1-p)^{N-{m}}$, where $p=\xi/L$.
In the limit of a wide disordered region ($L\to\infty$),
the product $pN$ remaining constant, the binomial law can be
approximated by a Poisson law  of parameter $\zeta=pN$: 
$\pi({m})\simeq e^{-\zeta}\,\zeta^{m}/m!$~. In
this limit the ${m}_i$'s are uncorrelated and the cumulative distribution
of $M$ is ${\mathscr F}_{\sss\!\!
  L}(M)=f(M)^B$ where $f(M)=\Gamma(M+1,\zeta)/M!$ is the
cumulative distribution function associated with the Poisson law 
[$\Gamma(x,\zeta)$ is the incomplete gamma function].

One can relate the SF critical velocity to the strength
$\Lambda_{\rm e}=M\,\lambda$ of the strongest effective peak by using 
the criterion for the critical velocity obtained in \cite{Leb01} for a
single peak: $\Lambda_{\rm e}=K(V_c/c_0)$, where
\begin{equation}
    K(z)=\frac{\sqrt{2}}{4z}
\left\{-8z^4-20z^2+1+(1+8z^2)^{3/2}\right\}^{1/2}\; .
  \label{delta_peak}
\end{equation}
The probability distribution of $V_c$ is then
\begin{equation}
P_L(V_c)=\frac{ K'(V_c/c_0)}{\lambda\,c_0}\;
 {\mathscr P}_{\sss\!\! L}(M=\Lambda_{\rm e}(V_c)/\lambda)\, ,
  \label{pvcdelta}
\end{equation}
where ${\mathscr P}_{\sss\!\! L}(M)=d{\mathscr F}_{\sss\!\! L}(M)/dM$.
\begin{figure}
  \includegraphics[width=0.95\linewidth]{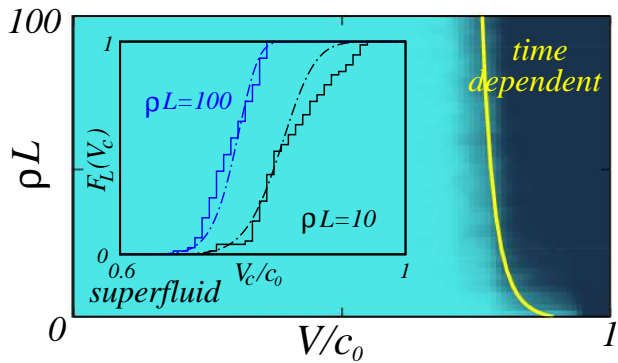}
  \caption{(Color online) Same as Fig.~\ref{diag_gauss} but for a
    random potential formed by a sequence of uncorrelated delta
    peaks ($\lambda=0.1$, $\rho\,\xi=0.1$).}
  \label{diag_delta1}
\end{figure}
Defining again the typical critical velocity as the median, the
boundary of the SF region corresponds to
\begin{equation}  \label{bound_delta}
  \frac{L(V_c)}{\xi} = 
\frac{ \ln 1/2}{\ln f(\Lambda_{\rm e}(V_c)/\lambda)} \; .
  \end{equation}
Figure \ref{diag_delta1} displays the stability phase
diagram and the cumulative distribution of the critical velocity. Here
also we find excellent agreement with the numerical results.

\section{Discusssion}

There are several experimental possibilities for studying 
the statistical properties of the boundary of the SF region.
The Gaussian disorder
with zero average may be implemented experimentally in the case of
micro-fabricated circuits, where the atoms are magnetically guided over a chip
\cite{chip}. Roughness and disorder in the circuits induce fluctuations along
the guide which are typically Lorentzian correlated, with a correlation length
$\ell_c$ which decreases with increasing distance between the guide and the chip
\cite{Pau05}.  However, the most common type of experimental disorder is the
so called speckle potential, generated by a laser beam passing through a
diffusing plate \cite{Clemspeck}. One of the most appropriate set ups seems to
be the one used in Ref.~\cite{Engels07}, where the critical velocity of a
trapped Bose-Einstein condensate has been probed by sweeping a
laser beam through it. The critical velocity was determined from measurements
of the amount of excitations related to the emission of solitons and linear
excitations.  Similar studies could be done by sweeping the laser beam through
a diffusive plate at constant velocity, thus creating a moving speckle
potential. Finally, the statistical properties of the SF breaking may also be
studied by simply adapting our calculations to the damping of dipole
oscillations, along the lines of the recent experiments of
Ref.~\cite{Hulet09}.

In conclusion we have stressed the link between the SF critical
velocity in presence of disorder and the statistics of extreme events.
We have developed simple models in two opposite situations where the
disorder is either very smooth or composed of point-like impurities.
In both cases the agreement between numerical simulations and our
analytical model is very good.  We note that, because of the mapping
of the statistical properties of the critical velocities to that of
extreme events of an uncorrelated sequence, in the limit $L\to\infty$
the distribution of $V_c$ tends to one of the universal distributions
of extreme value statistics \cite{extremevalues}.  A possible
extension of this work could be the application to fermionic
superfluids in the BCS regime.  In that case, the Landau critical
velocity is related to the pairing gap, and an equation similar to
(\ref{vc0}) can be explicitly written down \cite{Stri09}.

\begin{acknowledgments}We are grateful to S. Majumdar and J.
  Randon-Furling for fruitful discussions.  This work was supported by
  Grant No. ANR-08-BLAN-0165-01 and by the IFRAF Institute.
\end{acknowledgments}    \affiliation{Univ. Paris Sud, CNRS, Laboratoire de Physique
    Th\'eorique et Mod\`eles Statistiques, UMR8626, F-91405 Orsay}


\begin{thebibliography}{99}
\bibitem{rem1} Note that the time-dependent extension of
  Eq. (\ref{10}) in the case $\nu=1/2$ is delicate. However, its use
  in the stationary regime is perfectly legitimate.
\bibitem{rem2} $\gamma$ is obtained by comparing the exact extreme
  value statistics of the maxima a continuous disordered potential
  with the result of the discrete model. Note however that $\gamma$ is
  not a fitting parameter.
\end{thebibliography}
\end{document}